\documentclass{cupconf}
\usepackage{epsf}
\usepackage{psfig}

  \checkfont{eurm10}
  \iffontfound
    \IfFileExists{upmath.sty}
      {\typeout{^^JFound AMS Euler Roman fonts on the system,
                   using the 'upmath' package.^^J}%
       \usepackage{upmath}}
      {\typeout{^^JFound AMS Euler Roman fonts on the system, but you
                   dont seem to have the}%
       \typeout{'upmath' package installed. cupconf.cls can take advantage
                 of these fonts,^^Jif you use 'upmath' package.^^J}%
      }
  \else
  \fi

  \checkfont{msam10}
  \iffontfound
    \IfFileExists{amssymb.sty}
      {\typeout{^^JFound AMS Symbol fonts on the system, using the
                'amssymb' package.^^J}%
       \usepackage{amssymb}%
         \let\leq=\leqslant
         \let\geq=\geqslant
      }{}
  \fi

  \IfFileExists{amsbsy.sty}
    {\typeout{^^JFound the 'amsbsy' package on the system, using it.^^J}%
     \usepackage{amsbsy}}
    {}

\newsavebox{\astrutbox}
\sbox{\astrutbox}{\rule[-5pt]{0pt}{20pt}}

\newcommand\etal{\mbox{\textit{et al.}}}

\newcommand\Ha{H$\alpha$}
\newcommand\ha{\rm H\alpha}
\newcommand\hii{{\sc Hii}}
\newcommand\hi{{\sc Hi}}
\newcommand\msol{\rm\,M_\odot}
\newcommand\ergs{{\rm\,erg\,s^{-1}}}
\newcommand\kms{{\rm\,km\,s^{-1}}}

\newcommand\mmax{$m_{\rm max}$}
\newcommand\mup{$m_{\rm up}$}
\newcommand\nstar{$N_*$}
\newcommand\sfrcrit{SFR$_{\rm crit}$}
\newcommand\fwim{$f_{\rm WIM}$}
\newcommand\Sigha{$\Sigma_{\ha}$}
\newcommand\ApJ{{\it ApJ}\ }
\newcommand\aanda{{\it A\&A}\ }
\newcommand\AJ{{\it AJ}\ }
\newcommand\MNRAS{{\it MNRAS}\ }

\title[Various phenomena in fluid mechanics]{An interesting and seminal work
on various phenomena in fluid mechanics}
\title[Massive star feedback] 
{Massive Stars:  Feedback Effects \break in the Local Universe}

\author[M. S. Oey \& C. J. Clarke]%
{M.\ns S.\ns O\ls E\ls Y$^1$\ns
\and C.\ns J.\ns C\ls L\ls A\ls R\ls K\ls E$^2$}

\affiliation{$^1$Department of Astronomy, 830 Dennison Building, 
University of Michigan, Ann Arbor, MI\ \ \ 48109-1042, USA\\
$^2$Institute of Astronomy, University of Cambridge,
	Madingley Road, Cambridge\ \ \ CB3 0HA, UK}

\pubyear{2006}
\volume{538}
\pagerange{120--140}
\date{?? and in revised form ??}
\begin{document}
\maketitle

\begin{abstract}
Massive stars as a population are the source of various feedback
effects that critically impact the evolution of their host galaxies.
We examine parameterizations of the high-mass stellar population
and self-consistent parameterizations of the resulting feedback effects, including
mechanical feedback, radiative feedback, and chemical feedback, as we
understand them in the local universe.  To date, it appears that the
massive star population follows a simple power-law clustering law that
extends down to individual field massive stars, and the robust stellar IMF
appears to have a constant upper-mass limit.  These properties result in 
specific patterns in the \hii\ region luminosity function, and the
ionization of the diffuse, warm ionized medium.  The resulting
supernovae generate a population of superbubbles whose distributions
in size and expansion velocity are also described by simple power
laws, and from which a galaxy's porosity parameter is easily derived.
A critical star-formation threshold can then be estimated, above which
the escape of Lyman continuum photons, hot gas, and nucleosynthetic
products is predicted.  A first comparison with a large sample of \Ha\
observations of galaxies is broadly consistent with this prediction,
and suggests that ionizing photons are likely to escape from starburst galaxies.
The superbubble size distribution also offers a basis for a Simple
Inhomogeneous Model for galactic chemical evolution, which is
especially applicable to metal-poor systems and instantaneous
metallicity distributions.  This model offers an alternative
interpretation of the Galactic halo metallicity distribution and
emphasizes the relative importance of star-formation intensity, in
addition to age, in a system's evolution.  The fraction of
zero-metallicity, Population~III stars is easily predicted for any
such model.  We emphasize that all these phenomena can be modeled in a
simple, analytic framework over an extreme range in scale, offering
powerful tools for understanding the role of massive stars in the cosmos.
\end{abstract}

\section{Introduction}

Massive stars are of great interest because of their profound feedback
effects that alter the surrounding environment on local, global, and
cosmic scales.  Their radiative feedback causes ionization of neutral
gas; their supernova (SN) explosions drive mechanical feedback that
shock-heats gas to $\gtrsim 10^6$ K; and the nucleosynthesis processes
within these stars and their SNe produce most of the elements that are
tracers of past stellar populations.  In short, massive stars are one
of the principal drivers of galactic and cosmic evolution.

\section{The Massive Star Population}

By ``massive stars,'' we consider those stars having masses 
above, say, 10 $\msol$.  If we are to understand the global feedback
effects from massive stars, then it is important to understand their
properties as a {\it population}.  This population of stars is
characterized by its distribution in ($a$) mass, and ($b$) space.  

\subsection{The IMF and Upper-Mass Limit}

The stellar mass distribution is
parameterized by the familiar stellar initial mass function (IMF).
The massive star IMF has been evaluated many times for OB associations
and clusters in the Galaxy and Magellanic Clouds (e.g., Massey
2003), and it appears to be fairly robustly consistent with the
Salpeter (1955) slope: 
\begin{equation}\label{eq_IMF}
n(m)\ dm \propto m^{-2.35}\ dm
\end{equation}
where $n(m)$ is the number of stars in the mass range $m$ to $m+dm$.
Cruder evaluations of extragalactic massive star populations using
integrated colors and properties of galaxies support this result
(e.g., Elmegreen 2006; Fernandes \etal\ 2004; Bell \& de Jong 2001; Baldry \&
Glazebrook 2003).  An important possible exception may be the field
massive star IMF (see below).  Beware, however, that it is
difficult to disentangle effects of the IMF slope and upper-mass cutoff.

\begin{figure*}
\vspace*{0.5truein}
\hspace*{0.5truein}
\psfig{figure=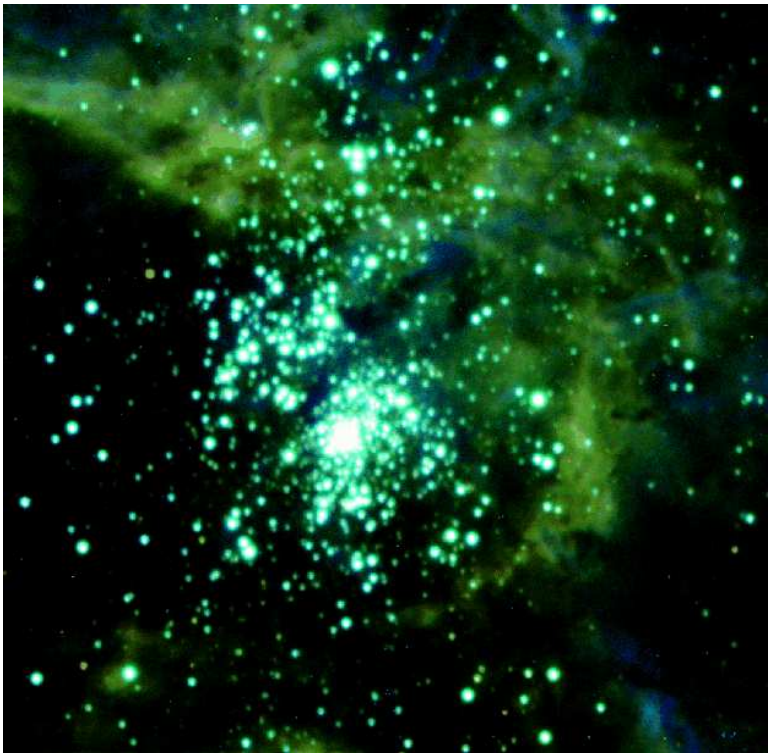,width=4.0truein}
\vspace*{0.3truein}
\caption{Three-color image of the R136a cluster in 30 Doradus, imaged
  with the NTT.  Red, green, and blue correspond to $V,\ B,$ and $U$,
  respectively.  (From Selman et al. 1999$a$.)
\label{oeyf_selman}}
\end{figure*}

While the slope of the upper IMF is fairly well-determined, the
upper-mass limit is less so.  There are theoretical considerations
supporting the existence of a stellar upper-mass limit based on
physical instability arguments (e.g. Ledoux 1941; Schwarzschild \&
H\"arm 1958; Stothers 1992), as well as limitations related to the high rate and
short timescales of accretion that are needed to overcome the
protostar's own radiation pressure (e.g. Larson \& Starrfield 1971;
Kahn 1974; Elmegreen \&  Lada 1977; Wolfire \& Cassinelli 1987; 
Bonnell, Bate, \& Zinnecker 1998).  These issues are also  
discussed elsewhere in these Proceedings (e.g., reviews
by Krumholz and Bonnell).  While the physical processes remain to be
fully understood, we can in the meantime empirically evaluate the
upper-mass limit, or lack thereof.

The IMF dictates that the highest-mass stars are the rarest, and so
the obvious place to search for such stars is in the richest clusters
that are young enough ($\lesssim$ 3 Myr) to preclude any having
exploded as SNe.  These rich, extremely young clusters are likewise rare
(see below), but we are fortunate that local examples do exist.  The
R136a cluster in the 30~Doradus star-forming complex in the Large
Magellanic Cloud (LMC; Figure~\ref{oeyf_selman}) is one such example.
Selman et al. (1999$b$)  
examined this region using extensive ground-based observations and
suggested that R136a exhibits an upper-mass cutoff around 150~$\msol$,
based on fairly qualitative arguments.  Massey \& Hunter (1998)
and Hunter et al. (1997) evaluated zero-age main sequence (ZAMS)
masses of hundreds of the most massive stellar candidates in R136a,
based on photometry and spectroscopic classifications from {\sl HST}.
Weidner \& Kroupa (2004) and Oey \& Clarke (2005) 
both examined the Massey \& Hunter statistics of these reported data
for R136a and independently concluded that this region quantitatively
demonstrates an upper-mass limit around $150 - 200\ \msol$.  Another
example of an extremely young and extremely rich cluster is the
Arches Cluster in the Galactic center environment.  It, too, exhibits
an upper-mass limit around $150 - 200\ \msol$ (Figer 2005).

But R136a and the Arches Cluster are only two specific regions.  Since
star formation presumably is a stochastic process, it may be that we
were simply extremely unlucky to have picked two clusters that both
happened to have rendered unusually low maximum stellar masses, even
though the physical mass limit may be much higher.  Furthermore, only
one or two more examples of rich clusters suitable for deterministic
evaluation of an upper-mass limit may be accessible for such detailed,
empirical, stellar mass analyses.  However, {\it if} the IMF indeed
behaves like a universal probability density function (PDF), then
we can also use a combined ensemble of the stellar contents of the
numerous, ordinary OB associations to evaluate an upper-mass limit.
Note that such treatment of the IMF as a PDF is indeed the 
conventional way in which it is usually considered.

\begin{figure*}
\psfig{figure=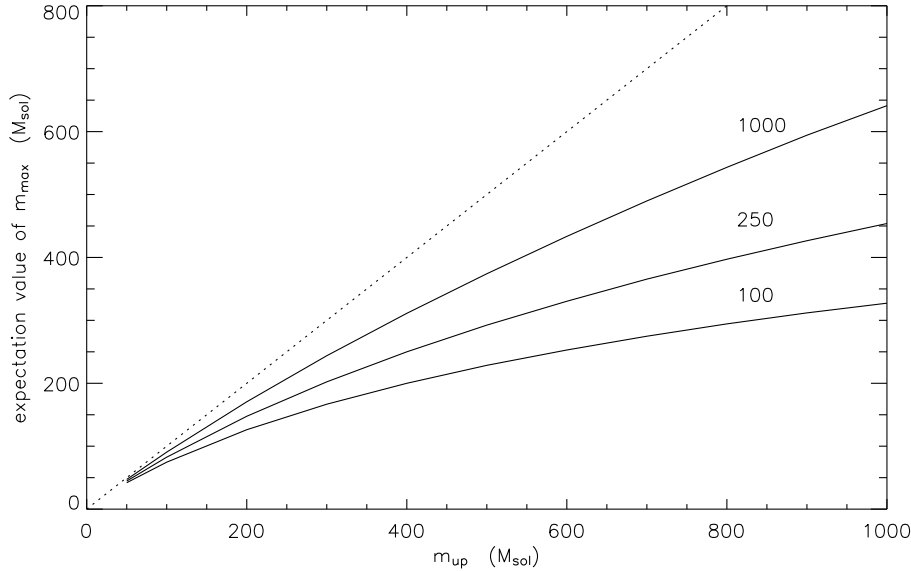,width=5.0truein}
\vspace*{-3.0truein}
\caption{Expectation value of the maximum  mass star in a cluster,
  given $N$ stars  having $m \geq 10\ \msol$, as a function of
  physical upper mass limit \mup.  (From Oey \& Clarke 2005.)
\label{oeyf_mmax}}
\end{figure*}

We can then assemble a large number of the youngest massive stars and
examine the form of the upper IMF.  For a Salpeter IMF, we can write
the expectation value of the maximum stellar mass \mmax\ for an
ensemble of \nstar\ stars having masses $m\geq 10\ \msol$, given a
physical upper-mass limit \mup: 
\begin{equation}\label{eq_mmax}
\langle m_{\rm max}\rangle = m_{\rm up} - 
   \int_{0}^{m_{\rm up}} \Biggl[\int_{0}^M \phi(m)\ dm\Biggr]^N \ dM \quad ,
\end{equation}
where $\phi(m)$ is the IMF.
Oey \& Clarke (2005) considered data for 8 OB associations in the
Galaxy and LMC in which the upper IMF was fully evaluated by Massey et
al. (1995), based on spectroscopic classifications, and which they
found to be $\leq 3$ Myr old.  Among these associations, there is a
total of 263 stars having $m\geq 10\ \msol$.  For a physical
\mup$=1000\ \msol$, equation~\ref{eq_mmax} predicts an expected
\mmax$> 450\ \msol$ (Figure~\ref{oeyf_mmax}), whereas the observed
maximum mass is $\lesssim 150\ \msol$.  Since R136a was
studied by the same group, we can include it in this uniform sample,
raising the total number to 913 stars having $m\geq 10\ \msol$.  For
the same parent IMF assumptions, Figure~\ref{oeyf_mmax} shows that the
predicted \mmax\ is now $\sim 600\ \msol$.  However, the 
observed maximum stellar mass in the entire ensemble is still only around
150~$\msol$, and so Figure~\ref{oeyf_mmax} implies a similar physical
upper-mass cutoff.  

Once again, we could be exceedingly unlucky in considering an
extraordinary sample of 9 associations, all of which happened to
render an unusually low \mmax.  Oey \& Clarke (2005) quantified the
likelihood of this occurrence as well, for physical \mup\ ranging from
$\infty$ to 120 $\msol$.  Only \mup$\sim 120 - 150\ \msol$ yielded
significant probabilities for the observed \mmax\ in the individual
clusters simultaneously, therefore clearly demonstrating an upper-mass
cutoff around those values.  Koen (2006) used an alternative statistical
analysis with these same data that confirms this result.

Clearly, this upper-mass limit is only demonstrated for this
particular sample of objects, and assuming that they are all pre-SN.
It seems significant, however, that the {\it same} \mup\ is found
across grossly varying environments encompassing the extreme
conditions near the Galactic Center for the Arches Cluster; the
highly-active, yet much less extreme conditions for R136a; and the
relatively unremarkable conditions for the OB associations.  These
findings suggest a universal upper-mass limit around 150
$\msol$ in the local universe.  Elmegreen (2000; 2006) makes similar
arguments in considering aggregrate stellar populations of galaxies.

\subsection{The Clustering Law}

With a fairly well-defined parameterization of the stellar mass distribution
for the upper IMF, we now turn to the spatial distribution of the
massive star population.  We can parameterize the space distribution
by defining a {\it clustering law}, $N(N_*)$, which describes the
distribution in $N_*$, the number of massive stars per cluster.
Over the last decade, it has become apparent that the clustering law
for young, massive clusters is robustly consistent with a power-law,
similar to the stellar IMF:
\begin{equation}\label{eq_clustering}
N(N_*)\ dN_* \propto N_*^{-2}\ dN_* \quad .
\end{equation}
This is equivalent to the initial cluster mass function and is
seen in a variety of environments, including starbursts and
populations of super star clusters (e.g., Meurer et al. 1995; Zhang \&
Fall 1999) and extrapolated from the globular cluster 
present-day mass function (e.g. Elmegreen \& Efremov 1997; Hunter et
al. 2003).

The upper-mass cutoff to the clustering law appears to vary in
different systems (see below), but no physical maximum has yet been
suggested.  In the opposite, low-mass extreme, the ``clusters'' reduce
to single, individual field massive stars.  With the definition of
``massive'' stars as above, ``field'' massive stars include both those
that are genuinely isolated from any other stars, if they exist, and
those that are the ``tip of the iceberg'' for a small group whose
remaining members all have $m$ less than would qualify for our
definition of a ``massive'' star.  

\begin{figure}
\hspace*{0.7truein}
\psfig{figure=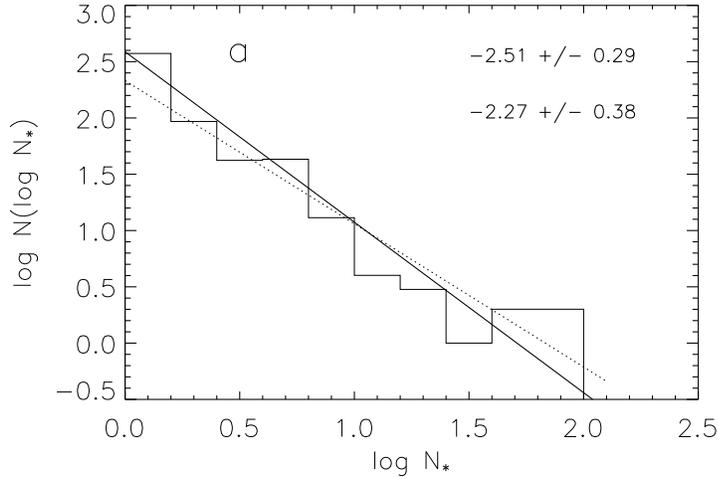,width=4.0truein}
\vspace*{-2.0truein}
\caption{SMC clustering law for OB star candidates, from Oey et
  al. (2004).  The solid line 
  shows a fit to the entire dataset, while the dotted line shows a fit
  omitting the field stars (first bin, at $\log N_*=0$).  These fitted
  slope values are shown, respectively.
\label{f_clustering}}
\end{figure}

What is the relationship between the field massive stars and those in
clusters?  Oey et al. (2004) studied a uniformly selected sample of
massive star candidates from $UBVR$ photometry (Massey et al. 2002) and
recalibrated FUV $B5$ (Parker et al. 1998) photometry across most of the Small
Magellanic Cloud (SMC) and determined the clustering law using a
friends-of-friends algorithm.  For that galaxy, they found that the
massive star clustering follows a smooth power law all the way down to
$N_* = 1$, and that the power-law exponent is consistent with
equation~\ref{eq_clustering} (Figure~\ref{f_clustering}). 

At face value, this suggests the co-existence of a universal IMF and
universal clustering law, of the respective forms in
equations~\ref{eq_IMF} and \ref{eq_clustering}.  However, we caution
that the field star IMF has been suggested to be significantly steeper
than in clusters, based on both observations (Massey 2002) and
theoretical arguments (Kroupa \& Weidner 2003).  If this is indeed
the case, then a flattening should be observed in the clustering law
near $N_* = 1$, since this implies fewer field stars.  Because this
flattening is not observed in Figure~\ref{f_clustering}, there must be
a corresponding steepening 
in the clustering law in this regime, in order to recover the smooth
power law that we see in the SMC data.  Thus reality may be somewhat
subtle and more complex than is seen at face value; Oey et al. (2004)
discuss this issue in more detail.  

Regardless of any underlying complexities, the resulting observed
clustering law in the SMC does show a smooth power law described by
equation~\ref{eq_clustering}.  If this applies generally, then we
can directly estimate the fraction of field massive stars as (Oey et
al. 2004):
\begin{equation}\label{eq_fieldfrac}
f_{\rm field} = \bigl(\ln N_{*,\rm max}+0.5772\bigr)^{-1} \quad , 
\end{equation}
where $N_{*,\rm max}$ is the number of massive stars in the richest
cluster of the ensemble.  Because of this dependence on
$N_{*,\rm max}$, we see that the field star fraction has a modest
inverse dependence on the total star formation rate of the system.
For typical star-forming galaxies, $f_{\rm field}\sim 20-25$\%.

The preceding thus describes a well-defined parameterization for
the massive star population in star-forming galaxies, given by
equations~\ref{eq_IMF} and \ref{eq_clustering}, and adopting a
stellar upper-mass cutoff \mup$=150 \msol$.  

\section{Radiative Feedback}

Now turning to the feedback effects from this massive star population,
we begin by considering the radiative feedback, which refers to the
photoionization caused by these stars.  There are two principal
effects:  the generation of ordinary \hii\ regions, and the generation
of the diffuse, nebular component of the interstellar medium (ISM).
The latter is usually referred to as the warm ionized medium (WIM), or
alternatively, diffuse ionized gas (DIG).

\subsection{The \hii\ Region Luminosity Function}

\begin{figure}
% \vspace{-3.5truein}
\hspace*{-0.5truein}
\psfig{figure=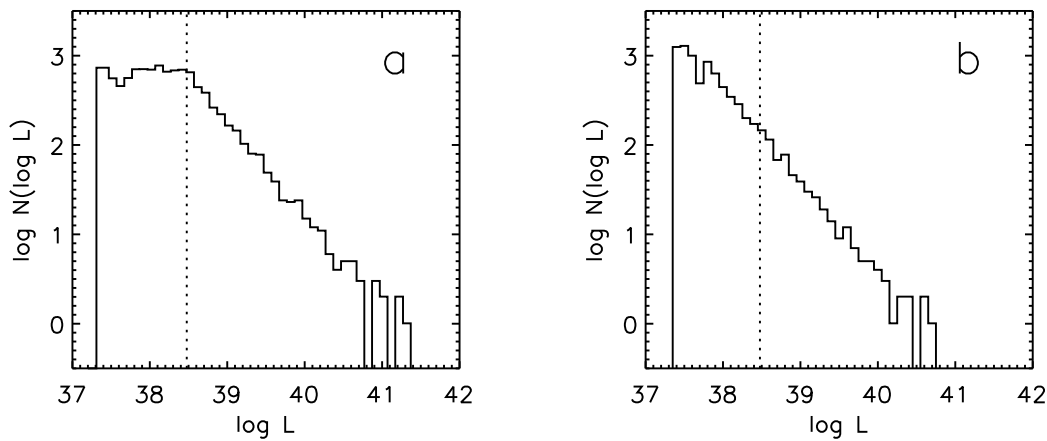,width=6.0truein}
\vspace*{-5.3truein}
\caption{Monte Carlo models of the zero-age \hii\ region luminosity
  function (panel $a$) and the same distribution evolved to 7 Myr
  (panel $b$).  The dotted lines show the \Ha\ luminosity associated
  with the most massive star in the IMF.  (From Oey \& Clarke 1998$a$.)
\label{f_hiilfev}}
\end{figure}

The \Ha\ luminosity is a direct probe of the cumulative massive star
population in an \hii\ region, and the clustering law naturally
results in a corresponding power-law luminosity function for the
classical \hii\ regions, which has been empirically examined in many
nearby galaxies.  However, the \hii\ region luminosity function
(\hii LF) is often seen to deviate from a smooth $N_*^{-2}$ power law
in the following ways:  (1) a two-slope form is often seen, with the
lower-luminosity population showing a shallower power-law slope; (2) 
inter-arm nebular populations in grand design spirals often show
shallower slopes than the arm populations; and (3) there is a
correlation with galaxy type, such that the early-type galaxies show
much steeper slopes than late types.  Oey \& Clarke (1998$a$) demonstrated,
using Monte Carlo models, that these variations are all fully
consistent with, and indeed expected, from the properties of the
massive star population described above.  One of the most important
effects is a flattening that is seen in the \hii LF at low
luminosities, that results from stochastic sampling of the stellar IMF
in this ``unsaturated'' regime (Figure~\ref{f_hiilfev}).  For luminous
nebulae generated by rich, ``saturated'' clusters that fully sample
the IMF through the upper-mass limit \mup, the \hii\ region luminosity
is directly proportional to $N_*$; 
whereas for sparse, ``unsaturated'' clusters, the resulting \hii\
region luminosity is subject to the specific stellar population.  The
stochastic flattening in the low-luminosity end of the \hii LF
quantitatively explains the observed effects (1) and (2) described
above.  The second of these, the observed steepening seen for
inter-arm \hii\ regions, can be explained if these show an evolved
slope as in Figure~\ref{f_hiilfev}$b$, while the zero-age population
in the arms corresponds to the model in Figure~\ref{f_hiilfev}$a$.
As a single-age population evolves, the entire distribution fades, and
the low-luminosity regime becomes dominated by evolved saturated objects.
Oey \& Clarke (1998$a$) demonstrate this effect in nearby
grand-design spirals, where the inter-arm regions are presumably a
more evolved population left behind in the wake of the spiral density
waves.

\begin{figure}
% \vspace{-3.5truein}
\hspace*{-0.5truein}
\psfig{figure=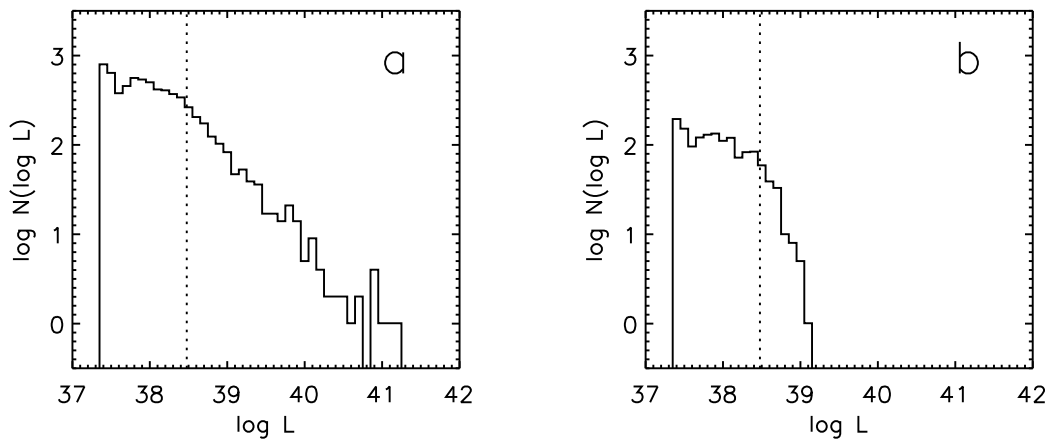,width=6.0truein}
\vspace*{-5.3truein}
\caption{Monte Carlo models of the \hii\ region luminosity function
  for continous star formation.  Panel $a$ shows the distribution for
  no upper-mass limit to the clusters, and panel $b$ shows the same,
  but imposing a limit of $N_*=10$.  (From Oey \& Clarke 1998$a$.)
\label{f_hiilfcut}}
\end{figure}

The third effect, the trend with galaxy Hubble type, can be explained
by a varying upper cutoff in $N_*$,  
while preserving the --2 power-law slope.  Figure~\ref{f_hiilfcut}$a$
shows a Monte Carlo model with no upper-mass limit to the clusters,
while Figure~\ref{f_hiilfcut}$b$ shows the same model \hii LF, but
imposing a limit of $N_* \leq 10$.  The former are qualitatively and
quantitatively similar to the observed \hii LF in late-type galaxies
(e.g., Kennicutt et al. 1989; Rand 1992; Banfi et al. 1993; Rozas et
al. 1996) while the former agree with the observations for Sa galaxies
(Caldwell et al. 1989)  Thus, all three of the observed patterns in
the \hii LF are fully consistent with, and indeed expected, for the
universal clustering law.  Oey \& Clarke (1998$a$) describe these
phenomena in detail. 

\subsection{The Diffuse, Warm Ionized Medium}

Classical \hii\ regions account for only about half of the total \Ha\
emission from star-forming galaxies (e.g., Walterbos 1998; Ferguson et
al. 1996).  The remaining half originates from the widespread, diffuse
WIM.  While the ionization of the WIM remains to be fully understood,
it generally thought also to originate from massive stars (e.g.,
Reynolds 1984).  Direct comparisons of 
the stellar populations in LMC OB associations with their associated
nebular luminosities suggests that up to 50\%, and in some cases,
more, of the ionizing radiation could escape from \hii\ regions (Oey
\& Kennicutt 1997; Voges et al. 2005).  Hoopes \& Walterbos (2003)
came to a similar conclusion based on FUV observations of M33 from
{\sl UIT}. 

The other half of WIM ionization can be accounted for by field stars,
assuming that the universal clustering law indeed extends to
individual massive stars representing the ``tip of the iceberg'' on
sparse clusters and groups, as found above for the SMC.
Equation~\ref{eq_fieldfrac} predicts that typically about 25\% of the
massive star population resides in the field as defined in this way, a
fraction confirmed empirically for the SMC (Oey et al. 2004).  Thus, field stars can
account for one-quarter of a galaxy's total \Ha\ luminosity.  For the
WIM constituting half of that total, then the field stars can ionize
about half again of the WIM. 

There are a few caveats; for example, other ionization processes are
implicated by apparent detailed ionization states observed in the WIM
(e.g. Reynolds \etal\ 1999; Rand 2000).  Also, the most recent hot star
atmosphere models (e.g. Martins et al. 2005; Repolust et al. 2005;
Smith et al. 2002) are suggesting softer ionizing fluxes which may
reduce the role of massive stars.  However, it seems clear that this
stellar population dominates production of the WIM.  

Radiative feedback to the IGM is also a topic of vital current interest,
and is discussed further below.

\section{Mechanical Feedback}

We now turn to the mechanical energy produced by the core-collapse
supernovae of the massive star population.  Given the short ($\leq 40$
Myr) lifetimes of these stars, the overwhelming majority remain in the
OB associations where they were born, and so the subsequent SNe are
spatially clustered.  Our universal $N_*^{-2}$ clustering law translates
directly into a mechanical luminosity function, that parameterizes the
kinetic energy for the ensemble of clusters.  Assuming that all SNe
yield the same kinetic energy, we can write the mechanical luminosity
function as 
\begin{equation}\label{eq_mechlf}
N(L)\ dL \propto L^{-2}\ dL \quad ,
\end{equation}
where $L$ is the ``mechanical luminosity'' or SN power expected from a
given cluster.  This makes the rough approximation that the discrete
SNe represent a continuous energy input over the 40 Myr timescale
(e.g., McCray \& Kafatos 1987).

\subsection{Superbubbles in the ISM}

The evolution of multi-SN superbubbles is given by simple, Sedov-like
relations (e.g., Weaver et al. 1977):  
\begin{equation}
R\propto \bigl (L/n \bigr)^{1/5}\ t^{2/5} \quad 
\end{equation}
\begin{equation}\label{eq_Pi}
P_i\propto L^{2/5}\ n^{3/5}\ t^{-2/5} \quad ,
\end{equation}
where $R$ and $P_i$ are, respectively, the superbubble radius and
interior pressure, and $n$ and $t$ are the ambient ISM density and
object age.  These relations assume adiabatic evolution, namely, that
the shells are pressure-driven by the shock-heated interior gas with
no thermal losses.

\begin{figure}
\vspace*{-2.0truein}
\hspace*{0.5truein}
\psfig{figure=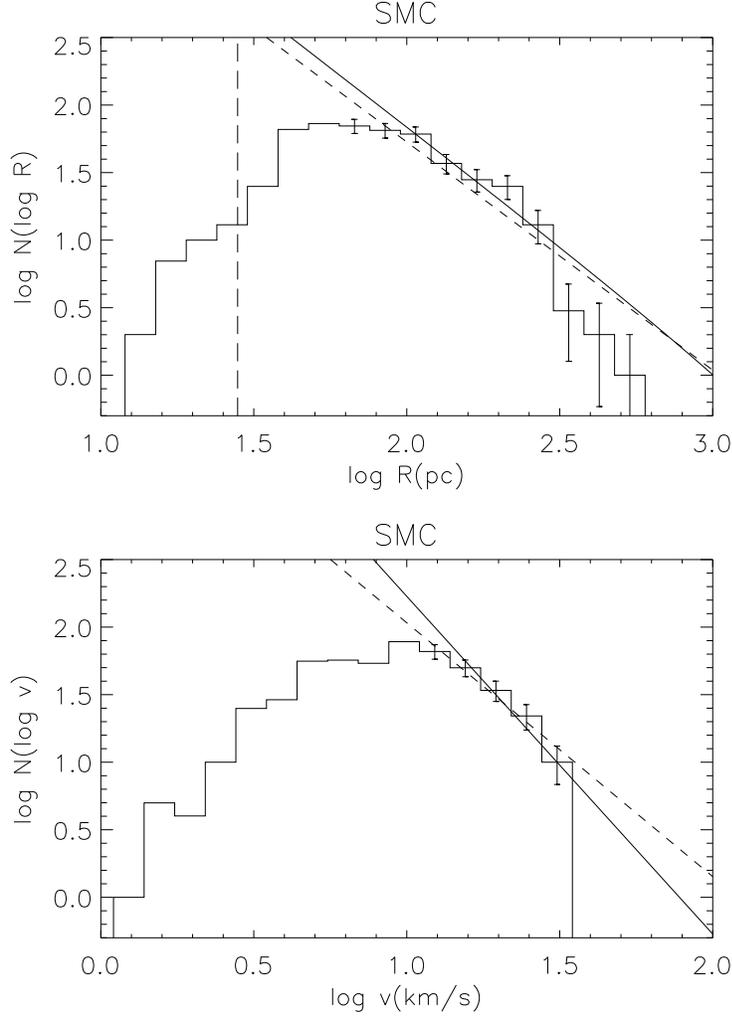,width=6.0truein}
% \vspace*{-3.0truein}
\caption{SMC \hi\ shell size distribution (top) and expansion
  velocity distribution (bottom) from the survey by
  Staveley-Smith \etal\ (1997).  The overplotted lines in the top
  panel show a power-law fit of $2.7\pm 0.6$ to the data (dashed), and
  a slope of $2.8\pm 0.4$ (solid) predicted from the observed \hii LF
  for this galaxy.  The spatial resolution of the \hi\ survey is shown
  by the vertical long-dashed line.  In the bottom panel, the solid line
  shows the predicted slope of --3.5, and the dashed line is a fit to
  the data, of $-2.9\pm 1.4$.  Note that only the high-velocity tail
  of the distribution corresponds to the expanding shells; the
  remainder are near the sound speed, and in pressure equilibrium with
  the ambient ISM.  (From Oey \& Clarke 1998$b$.)
\label{f_SMCshelldist}}
\end{figure}

For these simple analytic relations given by equations~\ref{eq_mechlf}
-- \ref{eq_Pi}, Oey \& Clarke (1997) derived global parameterizations of
superbubble populations, assuming continuous or burst creation
scenarios and that the shells stop growing when they are
pressure-confined by the ambient ISM.  For example, it is
straightforward to derive that the steady-state size distribution for
the mechanical luminosity function in equation~\ref{eq_mechlf} and a
continuous creation rate is,
\begin{equation}\label{eq_sizedist}
N(R)\propto R^{-3}\ dR \quad .
\end{equation}
At present, the only galaxies for which this prediction can be
reliably tested are the LMC and SMC, both of which have deep \hi\ surveys
and shell catalogs.  The top panel of Figure~\ref{f_SMCshelldist}
shows the size distribution for SMC \hi\ shell catalog (Staveley-Smith
et al. 1997), which is in excellent agreement with
the general prediction of equation~\ref{eq_sizedist} (Oey \& Clarke
1997).  The distribution in shell expansion velocities $v$, can be
similarly derived for the same population parameterizations (Oey \&
Clarke 1998$b$):
\begin{equation}\label{eq_velocitydist}
N(v)\ dv\propto v^{-7/2}\ dv \quad .
\end{equation}
The bottom panel of Figure~\ref{f_SMCshelldist} again shows agreement with this
prediction for the SMC shells, for those objects that are still
expanding (Oey \& Clarke 1998$b$).  On the other hand, the LMC shell
population (Kim et al. 1999) is different in nature.  Whereas the SMC
catalog has $> 500$ distinct \hi\ shells (Staveley-Smith et al. 1997),
the LMC catalog has only 126 coherent objects, in a survey with
similar instrumental sensitivities.  Since the LMC is larger and has a
substantially 
higher star formation rate than the SMC, the smaller shell population
is at first sight counter-intuitive.  However, as discussed below, the
LMC's star-formation rate appears to be high enough that the shells frequently
interact and merge, thereby losing their individual entities.  Our
predictions for the shell size distribution and other global
parameters cannot apply in such circumstances.

\subsection{The Threshold SFR for Feedback to the IGM}

Indeed, we can derive a threshold star-formation rate (SFR) above which we expect this
condition of shell interactions and ISM shredding.  The porosity
parameter $Q$ is a conventional way to parameterize the hot ($10^6$ K)
ionized medium (HIM) in galaxies.  $Q$ is essentially the filling
factor of this hot gas, and, since it is thought to originate from
shock-heating by SN explosions, $Q$ can be estimated as the total
volume of superbubbles relative to the relevant galaxy volume.  For
$Q>1$, the galaxy is generating more hot gas than it can contain, and
an outflow is predicted.  This also implies that the neutral ISM is shredded,
thereby allowing the escape of ionizing photons from the massive star
population.  Oey \& Clarke (1997) derived specific relations for $Q$,
for two- and three-dimensional situations, and Clarke \& Oey (2002) derived the
critical star-formation rate SFR$_{\rm crit}$ in general terms:
\begin{equation}\label{eq_sfrcrit}
{\rm SFR_{crit}} = 0.15\Biggl(\frac{M_{\rm ISM,10}
  \tilde{v}_{10}^2}{f_d}\Biggr ) \ \ \rm M_\odot\ yr^{-1} \quad ,
\end{equation}
where $M_{\rm ISM,10}$ is the ISM mass in units of 10$^{10}\ \msol$,
$\tilde{v}_{10}$ is the ISM thermal velocity in units of 10 $\kms$,
and $f_d$ is a geometric correction factor for disk galaxies.

For the Milky Way, Clarke \& Oey (2002) found that SFR$_{\rm crit} \sim 1\
\rm M_\odot\ yr^{-1}$, similar to our Galaxy's estimated star-formation
rate, and implying that it is close to this threshold.  They also
found that most local starburst galaxies might be expected to exceed
this criterion, since they often have smaller \sfrcrit\ but
larger star-formation rates in comparison to our Galaxy.  Lyman-break
galaxies should also show escaping ionizing radiation.  Note that the
\sfrcrit\ criterion is not based on an escape velocity, but 
rather an over-pressure.

\begin{figure*}
\psfig{figure=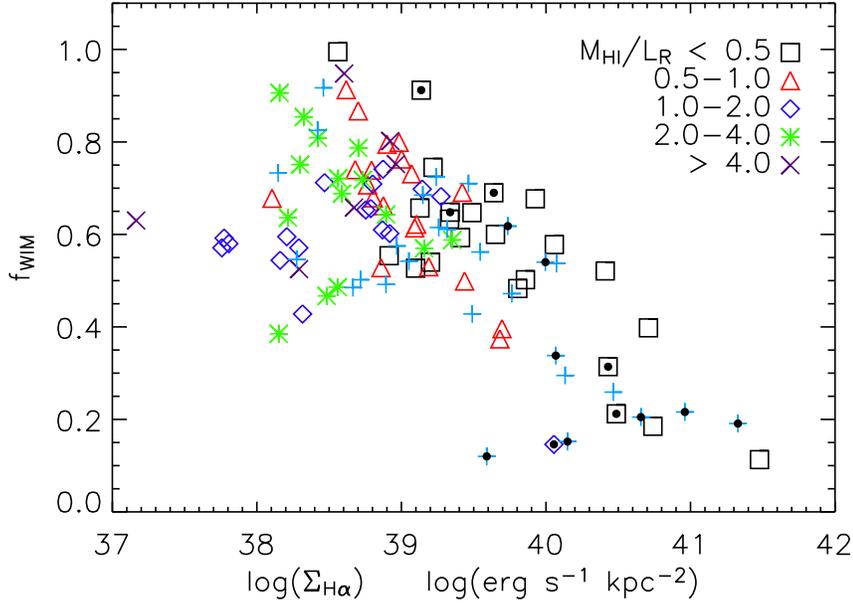,width=5.0truein}
% \vspace*{-2.0truein}
\caption{Fraction \fwim\ of diffuse emission relative to total \Ha\
  luminosity for 109 galaxies from the SINGG survey as a function of
  \Ha\ surface brightness \Sigha, computed within the \Ha\ half-light
  radius.  Symbols are assigned by \hi\ gas fraction $M_{\rm HI}/L_R$
  as shown, and black dots indicate galaxies dominated by  nuclear star
  formation.  (From Oey \etal\ 2006.)
% The solid curve shows a relation \fwim$\propto
%   L_{\ha}^{1/3}$ scaled arbitrarily (see text; 
\label{f_wim}}
\end{figure*}

\begin{figure*}
%\hspace*{-0.3truein}
%\vspace*{-1.0truein}
\psfig{figure=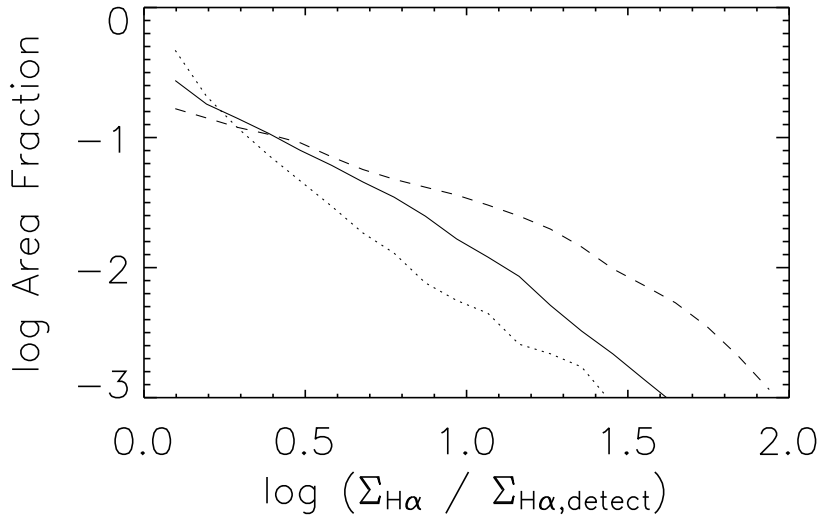,width=5.0truein}
%\vspace*{-0.5truein}
\vspace*{-3.3truein}
\caption{Co-added \Ha\ surface brightness distributions for the SINGG
  galaxies.  The dashed line shows ``starburst'' galaxies having
  log \Sigha\ $> 39.7$, the solid line shows galaxies with $38.7 <
  \log\Sigma_{\rm H\alpha} \leq 39.7$, and the dotted line shows galaxies
  with log \Sigha\ $\leq 38.7$; units of \Sigha\ are $\rm erg\ s^{-1}\
  kpc^{-2}$.  Only pixels with a signal above a 3$\sigma$ detection
  level are included.  (From Oey \etal\ 2006.)
\label{f_sbdist}}
\end{figure*}

A large sample of nearby galaxies that can be used to test this model
is the Survey for Ionization in Neutral-Gas Galaxies (SINGG; Meurer et
al. 2006), which is an \Ha\ survey of an optically-blind, \hi-selected
galaxy sample.  For the first dataset of 109 galaxies, we used the
HIIphot software of Thilker et al. (2000) to define the boundaries of
the classical \hii\ regions, assigning all remaining \Ha\ emission to
the WIM.  Figure~\ref{f_wim} shows the WIM fraction \fwim\ of the \Ha\
emission vs the \Ha\ surface brightness \Sigha\ for the sample.
Galaxies with the highest \Sigha\ within the star-forming disk show the
lowest \fwim.  We refer to 
galaxies having \Sigha\ $> 2.5\times 10^{39}\ \ergs\ \rm kpc^{-2}$ as
``starburst'' galaxies here, although Heckman (2005) defines
starbursts as much more intense systems having \Sigha $> 10^{41}\ \ergs\
\rm kpc^{-2}$.  Figure~\ref{f_sbdist} shows co-added \Ha\ surface
brightness distributions of galaxies in the SINGG sample, in three
bins of \Sigha.  It is again apparent that our starburst galaxies have
the flattest slope for the lowest surface brightnesses, also
demonstrating that they have the lowest \fwim.
% While this result has some dependence on the definitions
% for WIM vs \hii\ regions, the \Ha\ surface brightness distributions
% confirm that there is a real and significant effect.  We refer the
% reader to Oey et al. (2006) for a detailed discussion.

A possible cause for the lower \Ha\ diffuse fraction in starbursts
could be a lower fraction of ionizing field stars, as implied by
equation~\ref{eq_fieldfrac}.  If this is the dominant effect, then we
should similarly see it reflected in a relation between \fwim\ and the
total SFR as measured, for example, by the total \Ha\ luminosity.  However,
Oey et al. (2006) show that such an effect is not seen, and thus, the
lower fraction of field stars is not the dominant cause of the trend
in Figure~\ref{f_wim}.

We do note that local starbursts are expected to exceed the \sfrcrit\
threshold criterion for the escape of ionizing radiation
(equation~\ref{eq_sfrcrit}).  Figure~\ref{f_wim} shows that the
galaxies with the lowest \hi\ gas fractions, as measured by $M_{\rm
  HI}/L_R$, are those that most strongly exhibit the anti-correlation
in \fwim\ vs \Sigha, with $M_{\rm HI}$ corresponding to the \hi\
gas mass, and $L_R$ to the $R$-band luminosity.  The low \hi\ gas
fractions are consistent with a model in which there is not enough
neutral gas for the total ionizing luminosity, and some of the
radiation is lost from these galaxies.  Note that ionizing photons
could be lost through the shredded geometry of the ISM, as suggested
in the Clarke \& Oey (2002) model, or the ISM could simply be
fully ionized and density-bounded.  
% \begin{equation}
% f_{\rm WIM} \sim V_{\rm S,gal} - \sum{R_{{\sc Hii}}}^3 \quad ,
% \end{equation}
% where the first term on the right-hand side represents the galaxy
% total possible Str\"omgren volume, and the second term is the sum of
% all the normal \hii\ region volumes.  Thus we see that \fwim$\propto
% L_{\rm H\alpha}^{1/3}$.  Broadly taking all the galaxies to be the
% same size, we approximate t     The curve plotted in
% Figure~\ref{f_wim} shows a relation,
Since neutral gas is detected in
all of the galaxies, we favor the former model, but it may also be
possible that the star-forming disk is fully ionized and that the
unresolved \hi\ detections result from the outer regions of the
galaxies.  Oey et al. (2006) present a more complete discussion, in
which they show that a substantial fraction of the SINGG galaxies
exceed the \sfrcrit\ threshold for the escape of ionizing radiation,
including all of the starburst galaxies.  At the same time, they also
show that the relation between \fwim\ and \Sigha\ in
Figure~\ref{f_wim} is consistent with the simplest expectations for
density-bounding.  These results are strongly suggestive that ionizing
radiation is escaping from starburst galaxies through at least one of
these mechanisms.  Nevertheless, we caution
that several searches for Lyman continuum emission from
galaxies have yielded negative results (e.g., Heckman et
al. 2001; Leitherer et al. 1995); whereas more recently, the blue
compact dwarf galaxy Haro~11 does show a detection (Bergvall et
al. 2006), and at least two Lyman-break galaxies also show unambiguous
Lyman continuum emission (Shapley et al. 2006).  However, the positive
detections correspond to low escape fractions ($\lesssim 5$\%) of the
total ionizing radiation.  Further studies are
necessary to resolve these important and tantalizing issues regarding
the \Ha\ diffuse fraction and implied consequences. 

\section{Chemical Feedback}

The third feedback process is the nucleosynthesis by massive stars and
their core-collapse supernovae.  The element enrichment of the ISM in
galaxies and their environments drives the chemical evolution of
galaxies and the cosmos.  As another massive star feedback process,
chemical evolution can again be modeled with the same
parameterizations for the massive star population and mechanical
feedback processes.  The supernova activity heats the coronal
gas within superbubbles, which is the immediate medium into which the
nucleosynthetic products are injected.  The elements can mix and
disperse efficiently within this hot gas (Oey 2003; Tenorio-Tagle
1996), but their dispersal into cooler environments is a complex and
poorly understood process.  Pioneering simulations of the mixing and
dispersal process are only recently emerging (e.g., Balsara \& Kim
2005; de Avillez \& Mac Low 2003; see also Scalo \& Elmegreen 2004).

Oey (2000, 2003) introduced a rudimentary analytic model for galactic
chemical evolution that is based on the simple parameterizations
above.  The model assumes that the enrichment volume for an OB
association scales directly with the SN-driven superbubble volume,
with the former being the volume into which the SN products are
uniformly diluted.  The superbubble size distribution given by
equation~\ref{eq_sizedist} can therefore be used to derive the
relative metallicity distribution in the ISM for the ensemble of
massive star clusters given by equation~\ref{eq_clustering}.  Note
that the largest volumes dilute the products to the lowest
metallicities, and so Oey (2000) obtains:
\begin{equation}\label{eq_zdist}
N(Z)\propto Z^{-2}\ dZ \quad , \quad Z_{\rm min} < Z < Z_{\rm max} \quad .
\end{equation}
where $Z$ is the ISM metallicity, uniformly distributed within each
enrichment volume.  We impose the condition that no further dispersal
of the elements occurs beyond these individual volumes.  Thus, this
represents an extreme inhomogeneous, no-mixing model, which can be
contrasted to the opposite extreme of the pure, homogeneous Simple
Model (Schmidt 1963; Pagel \& Patchett 1975) that is used as the
standard reference for most studies of galactic chemical evolution. 

\begin{figure*}
\vspace*{0.3truein}
\hspace*{-0.5truein}
\psfig{figure=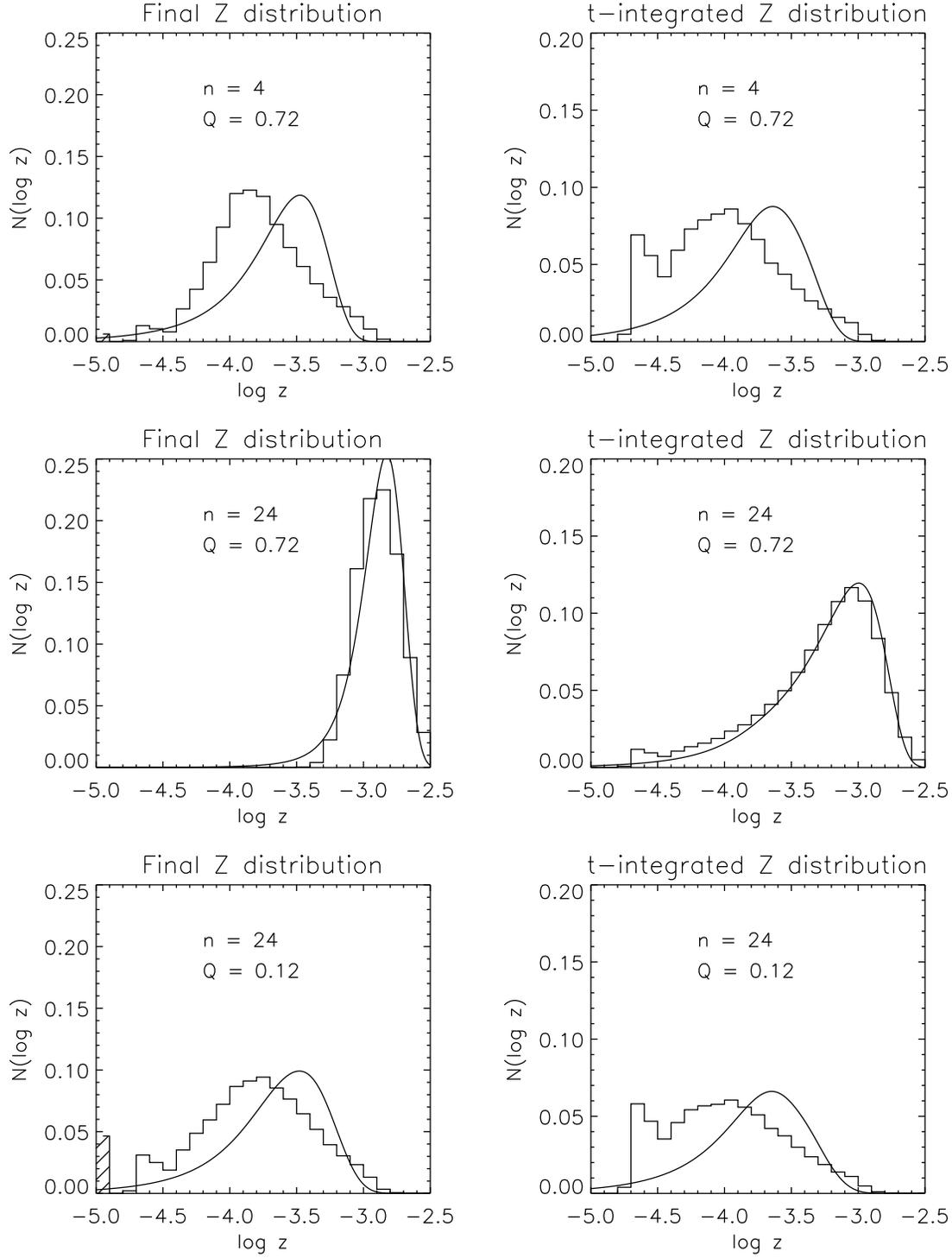,width=6.0truein}
% \vspace*{-2.0truein}
\caption{SIM models for different combinations of $n$ and $Q$ as
  shown, which correspond roughly to age and star-formation intensity,
  respectively.  Final, instantaneous MDFs are shown on the left side,
  and the corresponding time-integrated MDFs are on the right.  The
  SIM models are shown in the histograms, and analytic approximations,
  valid for large $nQ$, are shown by the curves. 
\label{f_nQ}}
\end{figure*}

Assuming that subsequent generations of massive stars generate the
same metallicity distribution (equation~\ref{eq_zdist}), we can model
the enrichment process by progressively summing the metallicities
as the volumes overlap.  Thus, after $n$
generations of star formation, the instantaneous metallicity
distribution function (MDF) is given by,
\begin{equation}\label{eq_mdfinst}
N_{\rm inst}(Z) = \sum_{j=1}^{n}\ P_j\ N_j(Z) \quad .
\end{equation}
where $N_j(Z)$ is the MDF for the ensemble of $j$ overlapping volumes,
which can be generated from the parent MDF given by
equation~\ref{eq_zdist}.  Note that any other form of the parent MDF can
also be used in place of equation~\ref{eq_zdist}, and also that
the Central Limit Theorem implies that $N_j(Z)$ approximates a
Gaussian distribution in the limit of large $j$.  $P_j$ is the
probability of obtaining $j$ overlapping regions, and is given by the
binomial distribution:
\begin{equation}
P_j = \Biggl({n \atop j}\Biggr)\ Q^j\ (1-Q)^{n-j} \ \ ,\quad
        1 \leq j \leq n \quad ,
\end{equation}
where $Q$ is the volume filling factor, that is again simply scaled 
from the porosity parameter considered above.

Equation~\ref{eq_mdfinst} therefore offers a Simple Inhomogeneous
Model (SIM), which is in the spirit of the homogeneous, Simple Model, but
which can be compared to observed {\it instantaneous} MDFs.  The SIM
is especially relevant to the most metal-poor systems, where
stochastic effects dominate the evolution.  It also emphasizes that
the star-formation intensity, not merely the simple age of the system,
is a major driver of the chemical evolution.  The parameter that
describes the evolutionary state in the SIM models is the
product $nQ$, which is also the mean of the binomial
distribution.  Figure~\ref{f_nQ} shows a series of SIM models for
different combinations of $n$ and $Q$, with instantaneous MDFs shown
on the left.  The right column shows corresponding models for
time-integrated MDFs (see below).  For the top and bottom models, the
product $nQ$ is the same, although the individual values of $n$ and
$Q$ differ; we see that the resulting MDFs are qualitatively and
quantitatively similar.  In contrast, the middle panels show a model
where the individual values of $n$ and $Q$ are the same as values in
the top or bottom, yet the product $nQ$ is much larger.  For this
model, the evolutionary state is much more evolved.
Thus, an old system with a low star-formation intensity
will have a similar MDF to a young system with a high star-formation
intensity.  This can straightforwardly explain the co-existence of
old, metal-rich systems like the Galactic bulge, and extremely
metal-poor systems like I~Zw~18 that also show old stellar
populations.  These two systems may have similar ages, but extremely
different time-integrated star-formation intensities, and they
therefore show very different metallicities and evolutionary state,
perhaps analogous to the middle and bottom models in Figure~\ref{f_nQ}.

\begin{figure*}
\hspace*{0.5truein}
\psfig{figure=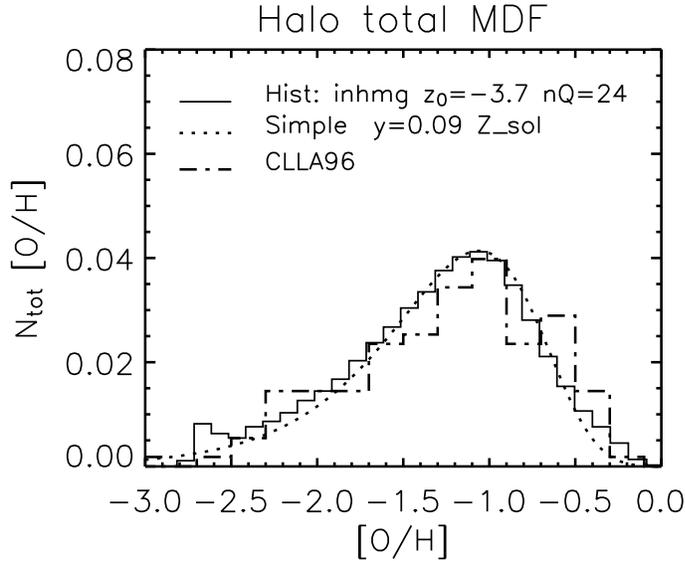,width=8.0truein}
% \vspace*{-2.0truein}
\caption{Halo metallicity distribution function converted to [O/H]
  from data by Carney \etal\ (1996, dot-dashed line).  A Simple
  Inhomogeneous Model is overplotted with the solid line, and a
  homogeneous Simple Model is overplotted with the dotted line.  The
  two models imply very different evolutionary states.  (Based
  on Oey 2003.)
\label{f_halo}}
\end{figure*}

\begin{figure*}
\vspace*{0.5truein}
\psfig{figure=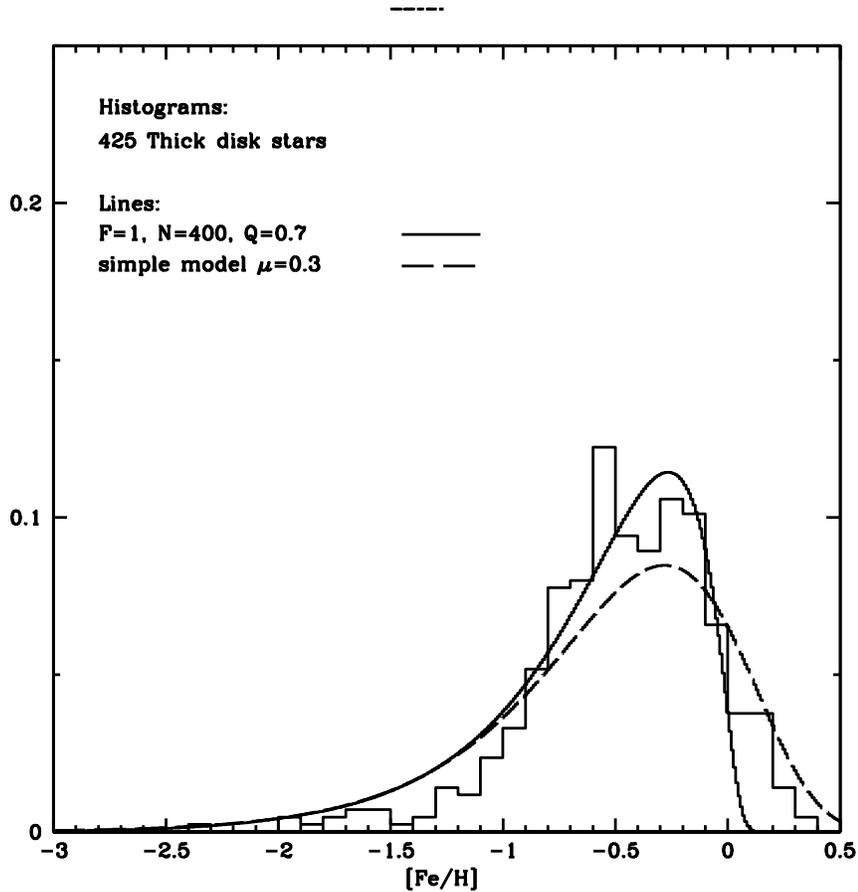,width=5.0truein}
% \vspace*{-2.0truein}
\caption{Metallicity distribution function for the Galactic thick
  disk from data by Nordstr\"om \etal\ (2004) and selection criteria
  of Bensby \etal\ (2003, 2005; solid histogram).  The overplotted
  curves show a Simple Inhomogeneous Model (solid) and a 
  Simple Model (dashed).  (From Bensby \& Oey 2006, in preparation.)
\label{f_td}}
\end{figure*}

Oey (2000, 2003) also derived a cumulative, time-integrated MDF for
all objects ever created out of the ISM whose enrichment proceeds in
this way, namely, long-lived stars:
\begin{equation}\label{eq_mdfcumu}
N_{\rm tot} =  \frac{1}{n}\ \sum_{j=1}^{n}\ \sum_{k=j}^{n}\ D_{k-1}\
        P_{j,k}\ N_j(Z)
\end{equation}
We now include a depletion factor $D_k$, which is the fraction of gas
remaining after $k$ generations of star formation.  Figure~\ref{f_halo}
compares the time-integrated SIM model (solid histogram) to the
Galactic halo MDF from the data of Carney et al. (1996; dot-dashed).
A homogeneous, Simple Model is also overplotted (dotted-line).  We see
that both models agree well with the data.  However, Oey (2003) shows
that the interpretations of these two models are extremely different:  the
Simple Model implies that the halo is a highly evolved system, because
the decrease in high-metallicity stars is caused by a lack of
remaining gas to form such stars.  In contrast, the SIM implies that
the halo is a relatively unevolved system, in which the decreasing
high-metallicity tail is still dominated by the form of the parent MDF
which is given by the $Z^{-2}$ distribution (equation~\ref{eq_zdist}).
Thus we see that further empirical constraints are necessary to
distinguish between these dramatically different possibilities.

Real systems should follow evolution that is intermediate between these
extremes described by the homogeneous Simple Model and inhomogeneous
SIM.  Figure~\ref{f_td} shows a preliminary MDF for the Galactic thick
disk.  The data correspond to F and G-dwarfs from the sample of
Nordstr\"om et al. (2004), selected according to kinematic critera of
Bensby et al. (2003, 2005).  We overplot both a Simple Model (dashed
line) and SIM model (solid line).  We see that
the thick disk data do in fact lie between the two models.  The only
exception is in the low-metallicity tail, where the so-called G-dwarf
Problem is seen:  there is a lack of the lowest-metallicity stars, in
comparison to both models, although the discrepancy is not as extreme
as for the Galactic thin disk.

Another useful feature of the SIM model is that it offers a
straightforward prediction for the fraction of zero-metallicity,
Population~III stars.  For any given SIM model, it is simply the
fraction of stars corresponding to $j=1$, which do not
overlap any preceding generations of contamination.  Thus,
\begin{equation}\label{eq_fiii}
F_{\rm III} = \sum_{k=1}^n\ D_{k-1} P_{1,k} \biggl / \
        \sum_{j=1}^n \sum_{k=j}^n D_{k-1} P_{j,k} \quad .
\end{equation}
For the Galactic halo SIM model shown in Figure~\ref{f_halo}, $F_{\rm
III} = 2\times 10^{-2}$.  As discussed by Oey (2003), this is two
orders of magnitude below the empirical upper limit of $2\times
10^{-4}$, demonstrating that the halo also shows a G-dwarf Problem, as
also found by others (e.g., Prantzos 2003).  Thus we see the power of these
simple analytic models in raising fundamental issues regarding the
formation and evolution of our Galaxy.

\section{Summary}

In summary, we see that empirical evidence thus far supports
simple parameterizations of the massive star population in terms of their
spatial distribution, given by the $N_*^{-2}$ clustering law
(equation~\ref{eq_clustering}), and mass 
distribution, given by a Salpeter (or similar) power-law IMF.  The
evidence is also suggestive, as described above, for a 
stellar upper-mass limit around 150 $\msol$, at least locally.  
The clustering law implies that about 10 -- 30\% of high-mass stars
are field stars (equation~\ref{eq_fieldfrac}), defined as those having no
other massive star siblings in the host cluster.  These
simple parameterizations of massive stars as a population have
powerful applications for analytically describing their feedback
effects.  

The ionizing radiation from these stars drives radiative feedback.
The clustering law directly results in the \hii\ region luminosity
function, which shows a similar --2  power-law exponent, provided that
the \hii\ regions have enough  ionizing stars to fully sample the
stellar IMF.  There is evidence that the upper limits to the \hii LF
vary across the Hubble Sequence, although the slope of the \hii LF
appears to be constant.  Star-forming galaxies possess an ISM component
at $10^4$ K, which also appears to be ionized by the high-mass stellar
population.  Typically, about half of the total \Ha\ emission from such
galaxies is observed to originate from ordinary \hii\ regions, while
the remaining half originates from the diffuse, warm ionized medium.
The diffuse WIM is likely ionized by both field stars and radiation
escaping from the \hii\ regions, in roughly equal parts.  However,
starburst galaxies show lower WIM fractions.  The origin of this
trend is unclear:  data from the SINGG survey are consistent with
predictions for escaping radiation based on
shredding of the ISM by mechanical feedback; on the other
hand, direct searches for Lyman continuum radiation from galaxies
have shown extremely low (~$\lesssim 5$\%) escape fractions.  

The global mechanical feedback from the massive star population is dominated
by their core-collapse SNe; stellar winds only play a significant role
for the youngest populations, in which the highest mass stars remain
unevolved and contribute the strongest stellar winds.  In a steady-state
scenario with constant global star-formation rate, the clustering law
implies a resulting superbubble size distribution with a dependence of
$R^{-3}$ (equation~\ref{eq_sizedist}) and expansion velocity
distribution dependence of $v^{-7/2}$ (equation~\ref{eq_velocitydist}).
These relations agree well for the \hi\ shell catalog for the SMC.  
From the superbubble size distribution, we
can derive a galaxy's porosity parameter or volume filling factor for
the superbubbles.  This is a standard convention for estimating the 
magnitude of the hot ($10^6$~K) component of the ISM, which is thought
to originate within the SN-heated interiors of the
superbubbles.  As the star-formation rate increases, the shells merge,
shred the cooler ISM, and generate more hot gas than the galaxy can
contain, thus driving a galactic outflow or superwind.  It is possible
to define a critical star-formation rate SFR$_{\rm crit}$, based on our
previous parameterizations and ISM properties
(equation~\ref{eq_sfrcrit}); we find that it is on the order of a few
$\msol\ \rm yr^{-1}$ for an $L^*$  galaxy.  We note that the LMC shows
only 1/4 the total number of coherent \hi\ shells compared to the SMC, 
despite its larger  size and star-formation rate.  This is consistent with
the prediction that it is near SFR$_{\rm crit}$, so that the shells
are merging and interacting.  Thus, SFR$_{\rm crit}$ represents a threshold
condition for the outflow of ionizing photons  through the shredded
ISM, as alluded to above, in addition to the outflow of hot gas and
the newly-produced SN products.  We note that this represents a
pressure-driven model, independent of a galaxy's escape velocity.

The clustering law and resulting superbubble size distribution also
offer a convenient framework for understanding the stochastic,
inhomogeneous progression of galactic chemical evolution.  Whereas the
standard, Simple Model is purely homogeneous at all times, the Simple
Inhomogeneous Model can make predictions for the {\it instantaneous}
metallicity distribution function at any snapshot in time
(equation~\ref{eq_mdfinst}).  It can also predict cumulative
MDFs (equation~\ref{eq_mdfcumu}), and so it can also
be applied to samples of long-lived stars.  The SIM is applicable, in
particular, to the most metal-poor conditions, and it agrees well with
the Galactic halo metallicity distribution.  This offers an
alternative interpretation of the halo as a relatively unevolved
population, in contrast to the highly-evolved status implied by the
Simple Model.  The SIM also offers an opposite extreme to
the homogeneous Simple Model, within which observations may be
bracketed.  In addition, it emphasizes that a system's evolutionary
status depends as much on the star-formation intensity, as on mere
age.  Finally, the SIM provides simple, straightforward
predictions for the fraction of zero-metallicity, Population~III stars
for any given model (equation~\ref{eq_fiii}).  Those for the Galactic halo confirm a
discrepancy of at least two orders of magnitude in the observed lack of
zero-metallicity stars compared to predicted fraction.

These diverse, massive-star feedback effects are all unified by the
same set of analytic parameterizations for this energetic stellar
population.  The observations are largely, and remarkably, consistent
across varying physical phenomena over an extreme range in scale and
age.  Thus, this simple, self-consistent framework offers powerful
tools and insight on the role of the massive star population in the
cosmos. 

\begin{acknowledgments}
MSO thanks the Symposium organizers for their hospitality and travel
support.  Some of this work was supported by the National Science
Foundation, grants AST-0448893 and AST-0448900; and the NASA
Astrophysics Data Program, grant NAG5-10768.
\end{acknowledgments}

% \clearpage

\end{document}